\documentclass[zpreprint,zbstpl]{./zeus_paper}
\usepackage[english]{babel}
\usepackage{verbatim}
\usepackage{url}
\chardef\usc=95
\chardef\til=126
\catcode`\@=11 
\DeclareRobustCommand\xdotspace{\futurelet\@let@token\@xdotspace}
\def\@xdotspace{%
  \ifx\@let@token.\else
  \ifx\@let@token\bgroup.\else
  \ifx\@let@token\egroup.\else
  \ifx\@let@token\/.\else
  \ifx\@let@token\ .\else
  \ifx\@let@token~.\else
  \ifx\@let@token!.\else
  \ifx\@let@token,.\else
  \ifx\@let@token:.\else
  \ifx\@let@token;.\else
  \ifx\@let@token?.\else
  \ifx\@let@token/.\else
  \ifx\@let@token'.\else
  \ifx\@let@token).\else
  \ifx\@let@token-.\else
  \ifx\@let@token\@xobeysp.\else
  \ifx\@let@token\space.\else
  \ifx\@let@token\@sptoken.\else
   .\space
   \fi\fi\fi\fi\fi\fi\fi\fi\fi\fi\fi\fi\fi\fi\fi\fi\fi\fi}
\catcode`\@=12 

\newcommand{\stru}[2]{%
   \relax\ifmmode\hbox{\vrule height#1 depth#2 width0pt}%
   \else\vrule height#1 depth#2 width0pt\fi}

\newcommand{\Ronum}[1]{\uppercase\expandafter{\romannumeral#1}}
\newcommand{\ronum}[1]{\expandafter{\romannumeral#1}}
\DeclareRobustCommand{\LaTeXZ}{%
  \LaTeX\kern-.05em4\kern-.1em
  {\raisebox{-0.2ex}{$\scriptstyle\text{ZEUS}$}}\xspace}

\DeclareMathAlphabet{\mathbf}{OT1}{cmr}{bx}{sl}
\newcommand{\eVdist}{\kern-0.06667em}

\newcommand{\mev}{{\,\text{Me}\eVdist\text{V\/}}}
\newcommand{\gev}{{\,\text{Ge}\eVdist\text{V\/}}}

\newcommand{\cm}{\,\text{cm}}

\newcommand{\rad}{\,\text{rad}}

\newcommand{\Tesla}{\,\text{T}}

\newcommand{\slashfrac}[2]{%
  \raisebox{0.5ex}{\ensuremath #1}\kern-0.12em/\kern-0.08em
  \raisebox{-.8ex}{\ensuremath #2}}

\newcommand{\sqr}[3]{%
    {\vcenter{\hrule height.#3ex\hbox{\vrule width.#2ex height#1ex
     \kern#1ex\vrule width.#3ex}\hrule height.#2ex}}}

\catcode`\@=11 
\newcommand{\parenbar}{\mathpalette\p@renb@r}
\def\p@renb@r#1#2{\vbox{%
  \ifx#1\scriptscriptstyle \dimen@.7em\dimen@ii.2em\else
  \ifx#1\scriptstyle \dimen@.8em\dimen@ii.25em\else
  \dimen@1em\dimen@ii.4em\fi\fi \offinterlineskip
  \ialign{\hfill##\hfill\cr
    \vbox{\hrule width\dimen@ii}\cr
    \noalign{\vskip-.3ex}%
    \hbox to\dimen@{$\mathchar300\hfil\mathchar301$}\cr
    \noalign{\vskip-.3ex}%
    $#1#2$\cr}}}
\catcode`\@=12 

\newcommand{\IP}{{\rm I$\kern-0.01667em$P}\xspace}

\mathchardef\qsm=63
\mathchardef\pls=43
\mathchardef\mns=512
\mathchardef\plm=518
\mathchardef\eql=61
\mathchardef\smallleft=300
\mathchardef\smallright=301
\mathchardef\les=316
\mathchardef\gre=318
\mathchardef\leq=532
\mathchardef\grq=533

\catcode`\@=11 
\newcounter{pict@width}
\newcounter{pict@height}
\newlength{\pict@scale}
\newcommand{\psfigadd}[4]{%
\setcounter{pict@width}{1*\ratio{#2+\pict@scale/2}{\pict@scale}}
\setcounter{pict@height}{1*\ratio{#3+\pict@scale/2}{\pict@scale}}
\setlength{\unitlength}{\pict@scale}
\hbox to #2{\hspace{-\fill}\begin{picture}(\thepict@width,\thepict@height)
\put(0,0){\psfig{figure=#1,width=#2,height=#3,clip=}}
\SetScale{0.283466457}
\SetWidth{1.763889}
{#4}
\end{picture}}
}
\newcounter{pict@widthfst}
\newcounter{pict@widthscd}
\newcounter{pict@widthtot}
\newcommand{\psfigaddtwo}[7]{%
\setcounter{pict@widthfst}{1*\ratio{#2+\pict@scale/2}{\pict@scale}}
\setcounter{pict@widthscd}{1*\ratio{#2+#4+\pict@scale/2}{\pict@scale}}
\setcounter{pict@widthtot}{1*\ratio{#2+#4+#6+\pict@scale/2}{\pict@scale}}
\setcounter{pict@height}{1*\ratio{#3+\pict@scale/2}{\pict@scale}}
\setlength{\unitlength}{\pict@scale}
\hbox{\hspace{-\fill}\begin{picture}(\thepict@widthtot,\thepict@height)
\put(0,0){\psfig{figure=#1,width=#2,height=#3,clip=}}
\put(\thepict@widthscd,0){\psfig{figure=#5,width=#6,height=#3,clip=}}
\SetScale{0.283466457}
\SetWidth{1.763889}
{#7}
\end{picture}}
}
\newcommand{\psfigror}[4]{%
\setcounter{pict@width}{1*\ratio{#2+\pict@scale/2}{\pict@scale}}
\setcounter{pict@height}{1*\ratio{#3+\pict@scale/2}{\pict@scale}}
\setlength{\unitlength}{\pict@scale}
\hbox{\begin{picture}(\thepict@width,\thepict@height)
\put(0,\thepict@height){\psfig{figure=#1,width=#3,height=#2,clip=,angle=270}}
\SetScale{0.283466457}
\SetWidth{1.763889}
{#4}
\end{picture}}
}
\newcommand{\psfigrol}[4]{%
\setcounter{pict@width}{1*\ratio{#2+\pict@scale/2}{\pict@scale}}
\setcounter{pict@height}{1*\ratio{#3+\pict@scale/2}{\pict@scale}}
\setlength{\unitlength}{\pict@scale}
\hbox{\begin{picture}(\thepict@width,\thepict@height)
\put(0,0){\psfig{figure=#1,width=#3,height=#2,clip=,angle=90}}
\SetScale{0.283466457}
\SetWidth{1.763889}
{#4}
\end{picture}}
}
\catcode`\@=12 
\newlength\listtextwidth

\catcode`\@=11 
\newlength{\@tabfninsert}
\newlength{\@tabfnwidth}
\newcommand{\tabfootnote}[2]{%
  \setlength{\@tabfninsert}{0.8em}
  \setlength{\@tabfnwidth}{\textwidth}
  \addtolength{\@tabfnwidth}{-\@tabfninsert}
  \addtolength{\@tabfnwidth}{-0.4em}
  \noindent\makebox[\@tabfninsert][r]{\footnotesize$^{#1}$\hfil}\hfill%
  \parbox[t]{\@tabfnwidth}{\footnotesize #2\hfill}}
\catcode`\@=12 

\def\citeCTD{{\cite{%
nim:a279:290,*npps:b32:181,*nim:a338:254%
}}\xspace}
\def\citeCAL{{\cite{%
nim:a309:77,*nim:a309:101,*nim:a321:356,*nim:a336:23%
}}\xspace}
\def\citeClose93{{\cite{%
pl:b319:291%
}}\xspace}

\def\citeWeinstein90{{\cite{%
prd:v41:2236%
}}\xspace}
\def\citeJaffe77{{\cite{%
prd:v15:267%
}}\xspace}
\def\citeAlford00{{\cite{%
np:b578:367%
}}\xspace}

\def\citedjango6{{\cite{%
cpc:81:381,*spi:www:djangoh11%
}}\xspace}
\def\citePDG{{\cite{%
epj:c15:1%
}}\xspace}

\def\citePDG02{{\cite{%
pr:d66:01%
}}\xspace}

\def\citeL3{{\cite{%
pl:b501:173%
}}\xspace}

\def\citeWAf1710{{\cite{%
pl:b453:305%
}}\xspace}

\def\citeBES96{{\cite{%
prl:v77:3959%
}}\xspace}

\def\citemark3{{\cite{%
prl:v56:107%
}}\xspace}

\def\citeH1Kaons{{\cite{%
np:b480:3,*zfp:c76:213%
}}\xspace}

\newcommand{\ksks}{$K_S^0 K_S^0$\xspace}
\newcommand{\Ks}{$K_S^0$\xspace}
\newcommand{\ffa}{$f_{2}(1270)/a_{2}^{0}(1320)$\xspace}
\newcommand{\ffb}{$f_{2}^{'}(1525)$\xspace}
\newcommand{\ffc}{$f_{0}(1710)$\xspace}

\begin{document}
\prepnum{{DESY--08--068}}
\date{June 2008}
\title{
Inclusive \boldmath$K^0_SK^0_S$ resonance production\\ 
in \boldmath$ep$ collisions at HERA}                                                       
                    
\author{ZEUS Collaboration}

\abstract{
Inclusive \ksks production in $ep$ collisions at HERA was studied 
with the ZEUS detector using an integrated luminosity of 0.5 fb$^{-1}$.
 Enhancements in the mass spectrum were observed and are attributed to
 the production of \ffa, \ffb and \ffc. Masses and widths were 
obtained using a fit which
 takes into account theoretical predictions based on SU(3) symmetry 
arguments, and are consistent with the PDG values.
The \ffc state, which has a mass consistent with a glueball candidate, was 
observed with a statistical significance of $5$ standard deviations. However, 
if this state is the same as that seen in $\gamma \gamma  \rightarrow K^0_S 
K^0_S$, it is unlikely to be a pure glueball state.
}

\makezeustitle
                                                                   
\pagenumbering{Roman}                                                                              

\begin{center}                                                                                     
{                      \Large  The ZEUS Collaboration              }                               
\end{center}                                                                                       
  S.~Chekanov,                                                                                     
  M.~Derrick,                                                                                      
  S.~Magill,                                                                                       
  B.~Musgrave,                                                                                     
  D.~Nicholass$^{   1}$,                                                                           
  \mbox{J.~Repond},                                                                                
  R.~Yoshida\\                                                                                     
 {\it Argonne National Laboratory, Argonne, Illinois 60439-4815, USA}~$^{n}$                       
\par \filbreak                                                                                     
  M.C.K.~Mattingly \\                                                                              
 {\it Andrews University, Berrien Springs, Michigan 49104-0380, USA}                               
\par \filbreak                                                                                     
  P.~Antonioli,                                                                                    
  G.~Bari,                                                                                         
  L.~Bellagamba,                                                                                   
  D.~Boscherini,                                                                                   
  A.~Bruni,                                                                                        
  G.~Bruni,                                                                                        
  F.~Cindolo,                                                                                      
  M.~Corradi,                                                                                      
\mbox{G.~Iacobucci},                                                                               
  A.~Margotti,                                                                                     
  R.~Nania,                                                                                        
  A.~Polini\\                                                                                      
  {\it INFN Bologna, Bologna, Italy}~$^{e}$
% ...
\newpage                                                                                           
\enlargethispage{5cm}                                                                              
$^{\    1}$ also affiliated with University College London, UK \\                                  
% ...
$^{\    5}$ now at University of Bonn, Germany \\                                                  
% ...
%
\newpage   
                                                           %
                                                           %
\begin{tabular}[h]{rp{14cm}}                                                                       
$^{a}$ &  supported by the Natural Sciences and Engineering Research Council of Canada (NSERC) \\  
$^{b}$ &  supported by the German Federal Ministry for Education and Research (BMBF), under        
          contract numbers 05 HZ6PDA, 05 HZ6GUA, 05 HZ6VFA and 05 HZ4KHA\\                         
$^{c}$ &  supported in part by the MINERVA Gesellschaft f\"ur Forschung GmbH, the Israel Science   
          Foundation (grant no. 293/02-11.2) and the U.S.-Israel Binational Science Foundation \\  
$^{d}$ &  supported by the Israel Science Foundation\\                                             
$^{e}$ &  supported by the Italian National Institute for Nuclear Physics (INFN) \\                
$^{f}$ &  supported by the Japanese Ministry of Education, Culture, Sports, Science and Technology 
          (MEXT) and its grants for Scientific Research\\                                          
$^{g}$ &  supported by the Korean Ministry of Education and Korea Science and Engineering          
          Foundation\\                                                                             
$^{h}$ &  supported by the Netherlands Foundation for Research on Matter (FOM)\\                   
$^{i}$ &  supported by the Polish State Committee for Scientific Research, project no.             
          DESY/256/2006 - 154/DES/2006/03\\                                                        
$^{j}$ &  partially supported by the German Federal Ministry for Education and Research (BMBF)\\   
$^{k}$ &  supported by RF Presidential grant N 8122.2006.2 for the leading                         
          scientific schools and by the Russian Ministry of Education and Science through its      
          grant for Scientific Research on High Energy Physics\\                                   
$^{l}$ &  supported by the Spanish Ministry of Education and Science through funds provided by     
          CICYT\\                                                                                  
$^{m}$ &  supported by the Science and Technology Facilities Council, UK\\                         
$^{n}$ &  supported by the US Department of Energy\\                                               
$^{o}$ &  supported by the US National Science Foundation. Any opinion,                            
findings and conclusions or recommendations expressed in this material                             
are those of the authors and do not necessarily reflect the views of the                           
National Science Foundation.\\                                                                     
$^{p}$ &  supported by the Polish Ministry of Science and Higher Education                         
as a scientific project (2006-2008)\\                                                              
$^{q}$ &  supported by FNRS and its associated funds (IISN and FRIA) and by an Inter-University    
          Attraction Poles Programme subsidised by the Belgian Federal Science Policy Office\\     
$^{r}$ &  supported by the Malaysian Ministry of Science, Technology and                           
Innovation/Akademi Sains Malaysia grant SAGA 66-02-03-0048\\                                       
\end{tabular}                                                                                      

\newpage

\pagenumbering{arabic}
\pagestyle{plain}

\section{Introduction}
\label{sec-int}

The existence of glueballs is predicted by QCD.
The lightest glueball is expected to have quantum numbers $J^{PC}=0^{++}$ 
and a mass in the range 1550--1750 MeV~\cite{Yao:2006px}.
Thus, it can mix with $q\overline{q}$ states
from the scalar meson nonet, which have
$I=0$ and similar masses.
Four states with $J^{PC}=0^{++}$ and $I=0$ are
established \cite{Yao:2006px}: $f_0(980)$, $f_0(1370)$, $f_0(1500)$ and $f_0(1710)$,
but only two states can fit into the nonet.
In the literature, the
state \ffc is  frequently considered to be a state
with a possible glueball or tetraquark composition~\cite{Klempt:2007cp,*Oller}.
However, its partonic content has yet to be established.

The ZEUS Collaboration previously observed~\cite{zeusKK} indications of two states,
\ffb and \ffc, decaying to \ksks final states in inclusive
deep inelastic scattering (DIS) events.
The statistical significance of the observation did not exceed three standard deviations.
The state in the 1700 MeV mass region had a mass consistent with that of the
\ffc;
however, its width was significantly narrower than that quoted by the Particle Data Group
(PDG)\cite{Yao:2006px}.

The results presented here correspond to the full HERA luminosity of $0.5$ fb$^{-1}$ and supersede the earlier ZEUS results. 
The measurement of the \ksks  final
states is presented in a kinematic region of $ep$ collisions dominated by photoproduction with
exchanged photon virtuality, $Q^2$,  below 1 $\gev^2$.
The data allow the reconstruction of the \ksks final states with much larger statistics than previously used.

\section{Experimental set-up}
\label{sec-exp}

The data were collected between 1996 and 2007 at the electron-proton collider HERA using the ZEUS detector.
During this period HERA operated with electrons or positrons\footnote{Here and in the following, the term ``electron" denotes generically
both the electron ($e^-$) and the positron ($e^+$).}
of energy $E_e=27.5$~GeV and protons initially with an energy of $820$~GeV and, after 1997, with
$920$~GeV.

A detailed description of the ZEUS detector can be found 
elsewhere~\cite{zeus:1993:bluebook}. Charged particles were tracked in the central tracking detector~\citeCTD 
which operated in a magnetic field of $1.43\Tesla$ provided by a thin 
superconducting solenoid. Before the 2004--2007 running period, the ZEUS tracking system
was upgraded with a silicon Micro Vertex Detector (MVD)~\cite{mvd2}. The high-resolution uranium--scintillator calorimeter (CAL)~\citeCAL consisted 
of three parts: the forward, the barrel and the rear calorimeters.

\section{Event sample}
\label{sec:data}

A three-level trigger system~\cite{zeus:1993:bluebook,Allfrey:2007zz} was used to select events online.
No explicit trigger requirement was applied for selecting \ksks events.
The photoproduction sample is dominated by events triggered by
a low jet transverse energy, $E_T$, requirement ($E_T>6\gev$). Deep inelastic scattering events 
were triggered by
requiring an electron in the CAL.

Events were selected offline by requiring $\mid Z_{\mathrm{vtx}} \mid  <   50\cm$,
where $Z_{\mathrm{vtx}}$  is  the $Z-$coordinate
of the primary vertex position
determined from the tracks.
The average energy of the total hadronic system, $W$, 
of the selected events was $\approx 200\gev$. The data sample was dominated by photoproduction events with $Q^2<1\gev^2$.

\section{Reconstruction of \Ks candidates
\label{reco-ks}}

The \Ks mesons were identified via their charged-decay mode,
$K_S^0 \to\pi^{+}\pi^{-}$.  Both tracks from the same secondary decay vertex were assigned the mass of the
charged pion and the invariant mass, $M(\pi^+\pi^-)$, of each track
pair was calculated. The \Ks candidates were selected by requiring:

\begin{itemize}

\item
$M(e^{+}e^{-})\ge 50\mev$, where the electron mass was assigned
to each track, to eliminate tracks from photon conversions;

\item
$M(p\pi)\ge 1121\mev$, where the proton mass was assigned to the track
  with higher momentum, to eliminate $\Lambda$ and
  $\bar{\Lambda}$ contamination to the \Ks signal;

\item
$p_T ( K_S^0 )\ge 0.25\gev$ and $|\eta ( K_S^0 ) |\le 1.6$, where $p_T ( K_S^0 )$ is the transverse momentum and
$\eta ( K_S^0 )$ is the pseudorapidity;

\item
$\theta_{2D} <~0.12~\rad$ ($\theta_{3D} <~0.24~\rad$), where $\theta_{2D}$
($\theta_{3D}$) is the two (three) dimensional collinearity angle between the \Ks-candidate 
momentum vector and the vector defined by the interaction point and the
\Ks decay
vertex. For $\theta_{2D}$, the $XY$ plane was used.

\end{itemize}

The cuts on the collinearity
angles significantly reduced the non-\Ks background
in the data during the 2004--2007 period. These cuts were necessary due to 
the extra material introduced by the MVD.
After all these cuts, the decay length distribution of the resulting \Ks candidates peaked
at $\approx 2\cm$.

Events with at least two \Ks candidates were accepted for further analysis. More than two \Ks were allowed in one event,
unlike for the previously published result~\cite{zeusKK}, and all distinct combinations of \ksks were used. 
In the mass range of $481\le M(\pi^+\pi^-)\le 515\mev$ the number of \Ks candidates is $1258399$.

\begin{figure}[p]
\vfill
\begin{center}
\epsfig{file=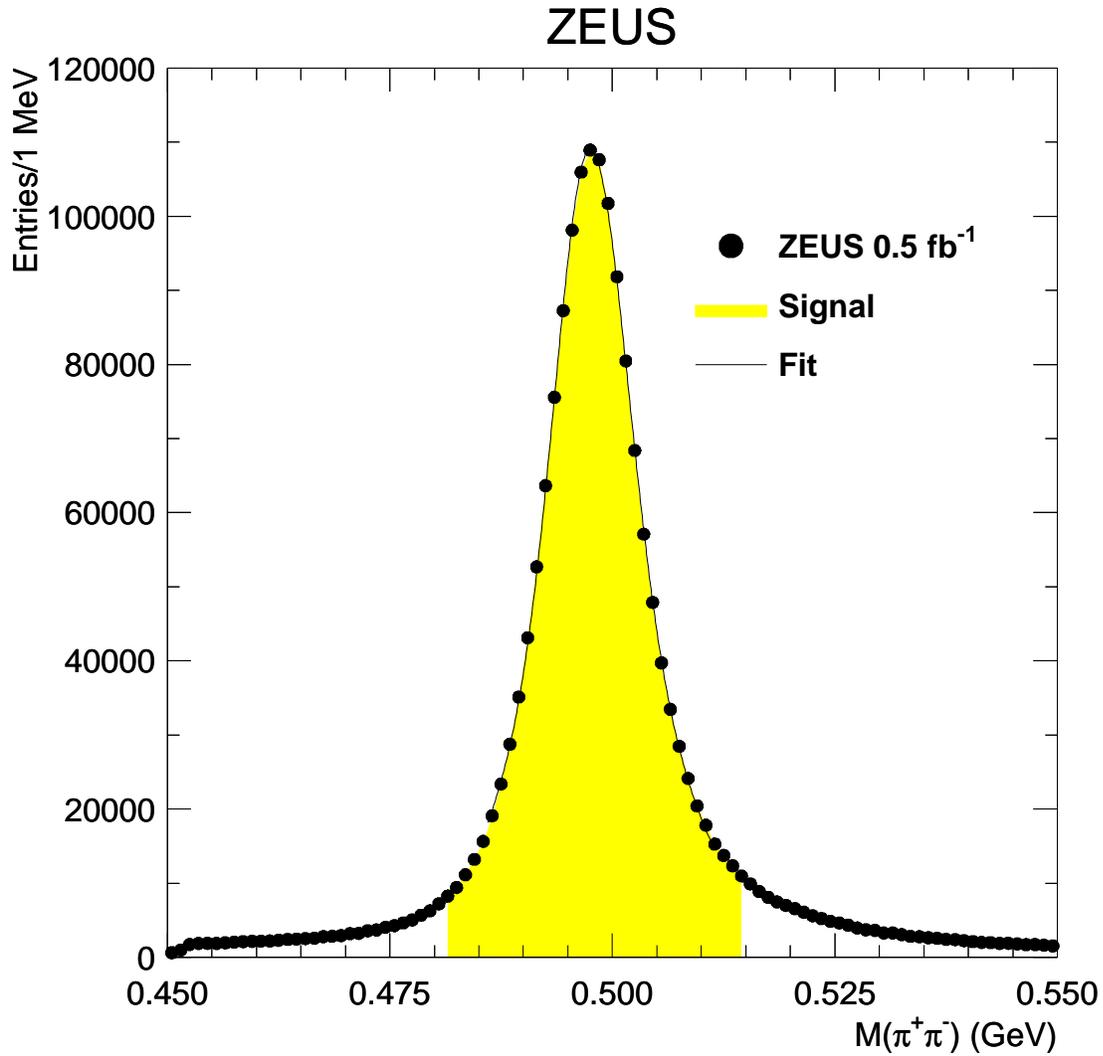,width=0.9\textwidth}
\end{center}
 \caption{The measured $\pi^+\pi^-$ invariant-mass distribution for events 
with at least two \Ks candidates (dots). The shaded area represents the signal window used for \ksks pair reconstruction. The fit performed (see text) is
displayed as a solid line. 
\label{fig:Ksmass} }
\vfill
\end{figure}

Figure~\ref{fig:Ksmass} shows the invariant-mass distribution of
\Ks candidates. A fit over the whole mass range including a first-order polynomial was used to estimate the background contribution at $\sim 8\%$. 
The central region was
fitted with two bifurcated Gaussian functions to determine the mass and width of the \Ks meson. 
For the HERA II data, corrections were applied to take into account the extra dead material introduced into the detector.
After the corrections, the mass and width of the \Ks were compatible with the PDG value and detector resolution, respectively.

\section{Results}
\label{sec-kk}

The \ksks invariant-mass distribution was reconstructed by combining two \Ks candidates
selected in the mass window $481\le M(\pi^+\pi^-)\le 515\mev$.
Tracks used for the \ksks pair reconstruction were required to be assigned uniquely to each \Ks in the \ksks pair.

\begin{figure}[p]
\vfill
\begin{center}
\epsfig{file=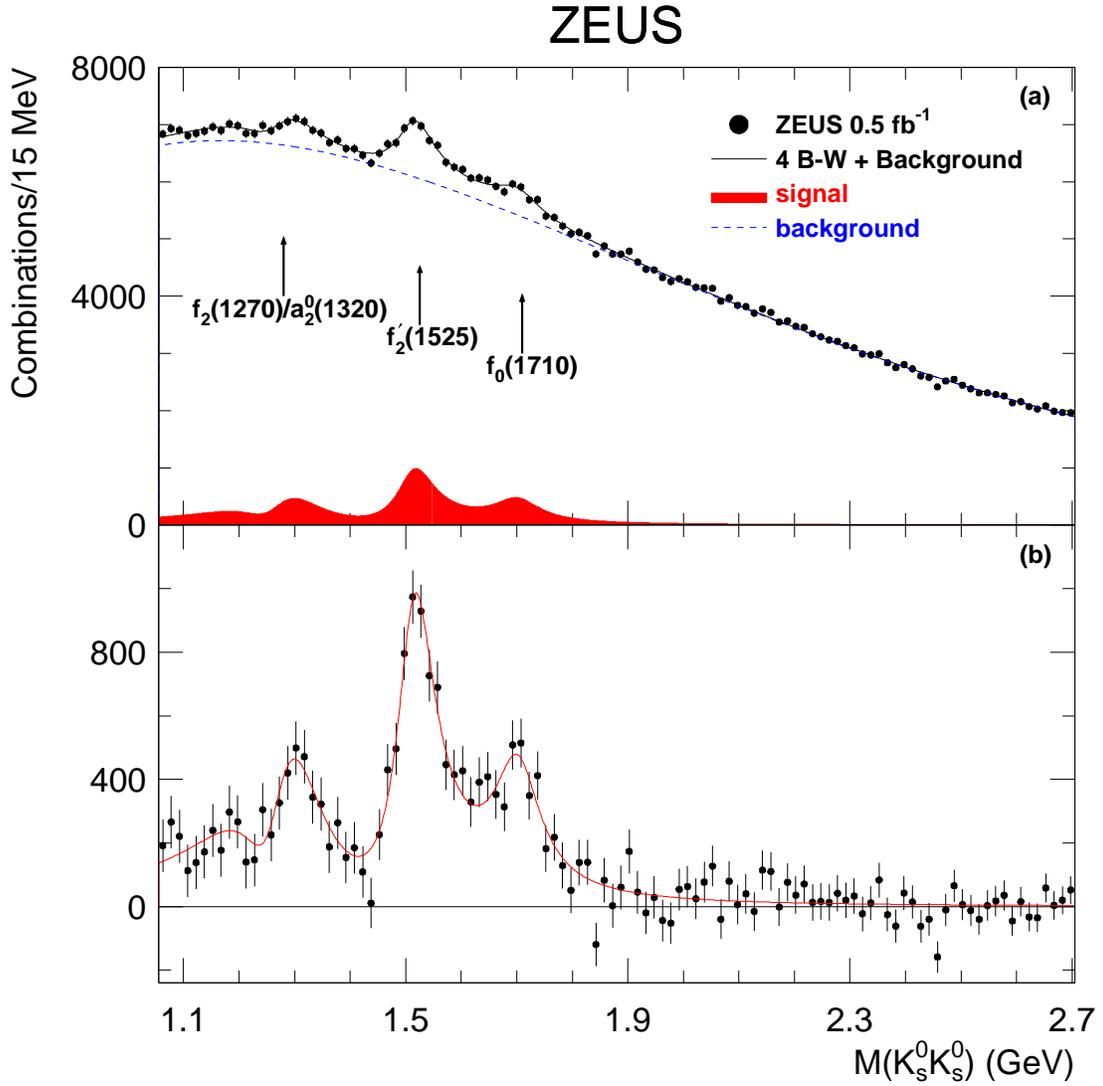,width=0.9\textwidth}
\end{center}
 \caption{(a) The measured \ksks invariant-mass spectrum (dots).  
The solid line is the result
of the fit described as fit 2 in the text (Eq.(3)) and the dashed line represents the background
function. (b) Background-subtracted \ksks
invariant-mass spectrum (dots); the result of the fit is shown as a solid line.
\label{fig:KKmass} }
\vfill
\end{figure}

Figure \ref{fig:KKmass}a shows the measured \ksks invariant-mass
spectrum.
Three peaks are seen at around 1300, 1500 and 1700 MeV. No state heavier than the \ffc was observed.
The invariant-mass distribution, \textit{m}, was fitted as a sum of
resonance production
 and a smoothly varying background $U(m)$. Each resonant amplitude, $R$, was given a
relativistic Breit-Wigner form~\cite{Althoff}:
\begin{equation}
BW(R) = \frac{M_R\sqrt{\Gamma_R}}{M_R^2-m^2-iM_R\Gamma_R}, \label{eq1}
\end{equation}
where $M_R$ and $\Gamma_R$ are the resonance mass and width, respectively.
The background function used was
\begin{equation}
U(m)~=~m^{A}\cdot \exp\left(-B
m\right)~, \label{eq2}
\end{equation}
where $A$ and $B$ are free parameters.
The \ksks mass resolution
is about $12\mev$ for the mass region below $1800\mev$ and
its impact on the extracted widths
is small compared to the expected widths
of the states~\cite{zeusKK}. Therefore, resolution effects were ignored in the fit.

Two types of fit, as performed for the reaction $\gamma \gamma \to$ \ksks by
the L3~\cite{Acciarri:2000ex} and TASSO~\cite{Althoff} Collaborations, respectively,
were tried, using Eqs. (1) and (2). Fit 1 is an incoherent sum of three
Breit-Wigner cross sections representing the \ffa, \ffb and
\ffc plus background. Fit 2 is motivated by SU(3) predictions~\cite{Faiman}. The decays of the tensor ($J^P=2^+$) mesons
$f_2(1270)$, $a^0_2(1320)$ and \ffb into the two pseudoscalar
($J^P=0^-$) mesons $K^0\bar{K^0}$ are related by SU(3) symmetry with a
specific interference pattern. The intensity is the modulus-squared of the
sum of these three amplitudes plus the incoherent addition of \ffc
and a non-resonant background. The predicted coefficients of the
$f_2(1270)$, $a^0_2(1320)$ and \ffb Breit-Wigner amplitudes for
an electromagnetic production process are, respectively, +5, -3 and +2~\cite{Faiman,priv:lipkin}.
This results in the fit function:
\begin{eqnarray}
F(m) &=& a\cdot|5\ BW(f_2(1270)) - 3\ BW(a_2^0(1320)) + 2\ BW(f_{2}^{'}(1525))|^2 +{}\nonumber\\
& + &
     b\cdot|BW(f_{0}(1710))|^2 + c\cdot U(m), \label{eq3}
\end{eqnarray} where $a$, $b$ and $c$ are free parameters.

All the resonance masses and widths were allowed to vary in the fits.
The results of the fits are shown in Table~1. The quality of both fits,
characterized by the $\chi^2$ per number of degrees of freedom (see Table~1), is good.
However, fit 2 describes the spectrum around the
\ffa region better and, unlike fit 1, reproduces the dip
between \ffa and \ffb. For this reason and,
based on the theoretical motivation, fit 2 is preferred and shown in Fig.~2. The
 background-subtracted mass spectrum is shown in Fig.~2b together with the fit.

The $a^0_2(1320)$ mass in fit 2 is below the PDG value~\cite{Yao:2006px}. A similar
shift, attributed to the destructive interference between $f_2(1270)$ and
$a^0_2(1320)$, was also seen in a study of resonance physics with $\gamma\gamma$ events~\cite{Faiman}. Fit 1 without
interference yields a narrow width for the combined
\ffa peak, as also seen by the L3 Collaboration~\cite{Acciarri:2000ex}. Fit 2 with
interference yields widths close to the PDG values for all observed
resonances. The fitted masses for \ffb and \ffc are
somewhat below the PDG values with uncertainties comparable with those of the PDG (Table~1). The quality of a fit without 
the \ffc resonance (not shown) yields $\chi^2$/ndf= 162/97
and is strongly disfavored.

The systematic uncertainties of the masses and widths of the resonances,
determined from the fit shown in Fig.~\ref{fig:KKmass}, were evaluated by
changing the selection cuts and the fitting procedure. Variations of minimum track $p_T$, track pseudorapidity range,
 track momenta by $\pm 0.1\%$, accepted $\pi^+\pi^-$
mass range around the \Ks peak and collinearity cuts were done. In addition a maximum
likelihood fit was used instead of the $\chi^2$ fit and event selection cuts
were varied. A check for the possible influence of the  $J^P=0^+$ state $f_0(1500)$
was carried out by including in the fit a Breit-Wigner amplitude of
this state interfering with the amplitude of the \ffc. The resulting
changes of the fitted values of the mass and the width of the
\ffc are included in the systematic uncertainties~\cite{thesis:czhou}.
 The largest systematic uncertainties were: fitting with fixed PDG mass and width on \ffb
affects the \ffc width by -19 MeV and the largest effect of varying the 
track momenta on the \ffc width is +7 MeV. The combined systematic 
uncertainties are
included in Table~1.

\begin{figure}[p]
\vfill
\begin{center}
\epsfig{file=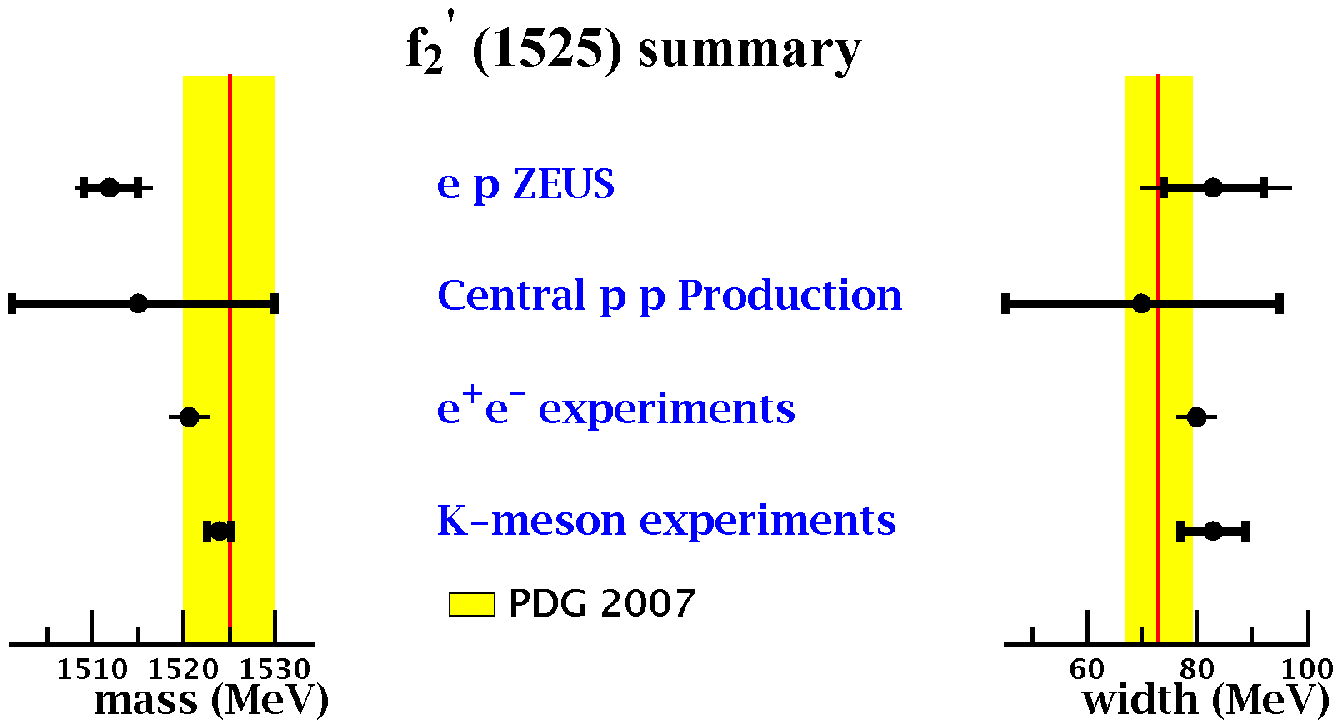,width=0.65\textwidth}

\vspace{1.0cm}

\epsfig{file=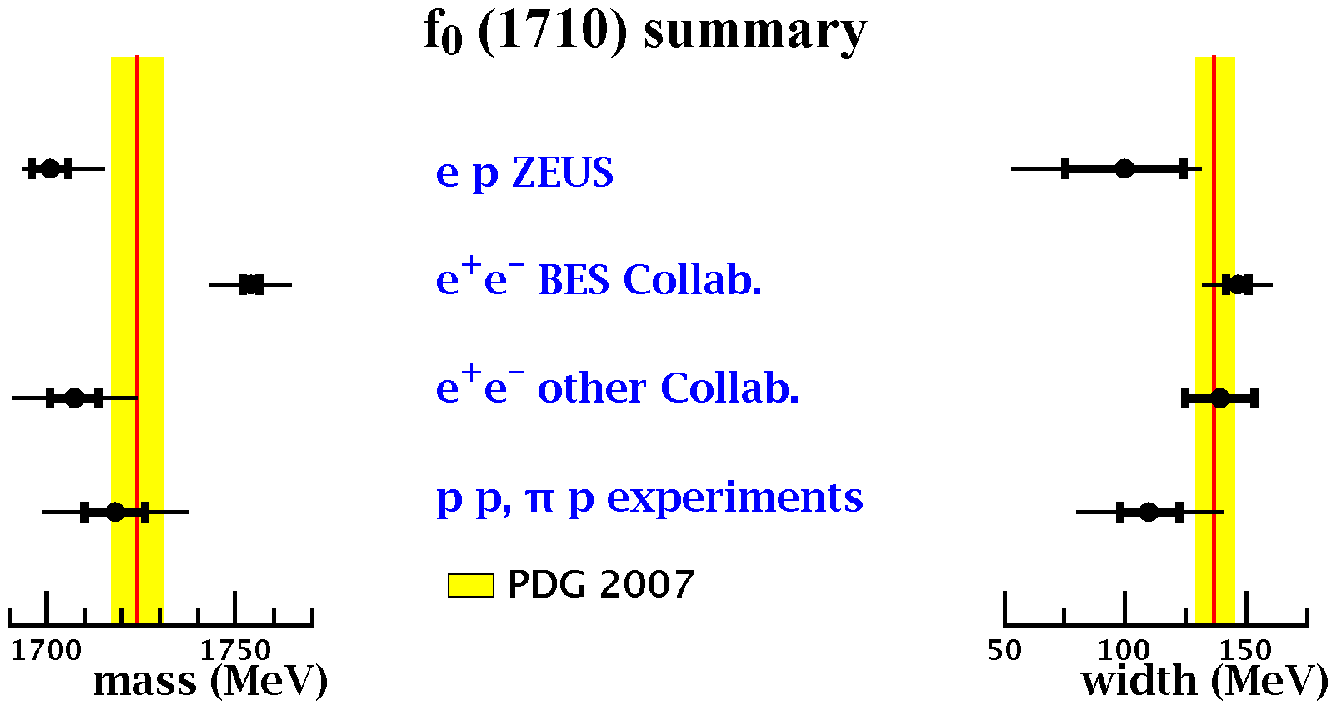,width=0.65\textwidth}

\end{center}
\caption{ 
Comparison of the present mass and width measurements of the \ffb and  \ffc states 
with other selected measurements~\protect\cite{Yao:2006px}. 
The bands show the PDG values and error estimates.    
\label{fig:sum1} }
\vfill
\end{figure}

The number of events in the \ffc resonance given by the fit is $4058 \pm 820$, which has a $5$ standard-deviation statistical significance. 
This is one of the best \ffc signals reported. 
This state is considered to be a glueball candidate~\cite{Klempt:2007cp,*Oller}. However, if it is the same as seen in $\gamma \gamma
 \rightarrow K^0_S K^0_S$~\cite{Althoff,Acciarri:2000ex}, it is unlikely to be a pure glueball state, 
since photons can couple in partonic level only to charged objects. 
Figure~\ref{fig:sum1} compares the results of this analysis with
other measurements from collider and fixed-target experiments. The \ffc mass as deduced from the quarkonium decays by the BES Collaboration is
significantly higher than the values given by all other experiments, including older $J/\psi$-decay analyses~\cite{Yao:2006px}.

\section{Conclusions}
\label{sec-con}

In conclusion, \ksks final states were studied
in $ep$ collisions at HERA with the ZEUS detector.
Three enhancements which correspond to \ffa, \ffb and \ffc were observed. No state heavier than the \ffc was observed.
 The states were fitted taking into account the interference pattern predicted by
SU(3) symmetry arguments.
The measured masses of the \ffb and \ffc states are somewhat below the world average, however, the widths are consistent with the
PDG values. The \ffc state, which has a mass consistent with a $J^{PC}=0^{++}$ glueball candidate, is observed with a $5$ standard-deviation
 statistical significance. However, if this state is the same as that seen in
 $\gamma \gamma
 \rightarrow K^0_S K^0_S$, it is unlikely to be a pure glueball state.

\section{Acknowledgements}
We thank the DESY directorate for their strong support and
encouragement. The special efforts of the HERA machine group
in the collection of the data used in this paper are
gratefully acknowledged. We are grateful for the support of the
DESY computing and network services. The design, construction
and installation of the ZEUS detector were made
possible by the ingenuity and effort of many people from DESY
and home institutes who are not listed as authors.
We also thank H.J. Lipkin for
valuable comments and advice.

\newpage

\providecommand{\etal}{et al.\xspace}
\providecommand{\coll}{Collaboration}
\catcode`\@=11
\def\@bibitem#1{%
\ifmc@bstsupport
  \mc@iftail{#1}%
    {;\newline\ignorespaces}%
    {\ifmc@first\else.\fi\orig@bibitem{#1}}
  \mc@firstfalse
\else
  \mc@iftail{#1}%
    {\ignorespaces}%
    {\orig@bibitem{#1}}%
\fi}%
\catcode`\@=12

\newpage

\begin{table}[!htb]

\begin{center}
\begin{tabular}{|c|c|c|c|c|c|c|}
\hline 
Fit & \multicolumn{2}{c|}{No interference} & \multicolumn{2}{c|}{Interference} &  \multicolumn{2}{c|}{} \\
\cline{1-5}
\raisebox{0.ex}[0.6ex]{$\chi^2/$ndf} & \multicolumn{2}{c|}{\raisebox{0.ex}[0.6ex]{96/95}} & 
\multicolumn{2}{c|}{\raisebox{0.ex}[0.6ex]{86/97}} & 
\multicolumn{2}{c|}{\raisebox{1.7ex}[0.7ex]{PDG 2007 Values}} \\[-0.1ex]
\hline
in MeV & Mass & Width & Mass & Width & Mass & Width \\
\hline
$f_2(1270)$ & & & $1268\pm 10$ &  $176\pm 17$ &  $1275.4\pm 1.1$ &  $185.2^{+3.1}_{-2.5}$\\
\cline{1-1}\cline{4-7}
\raisebox{0.ex}{$a_2^0(1320)$} & \raisebox{1.7ex}[0.8ex]{$1304\pm 6$} & \raisebox{1.7ex}[0.8ex]{$61\pm 11$} & \raisebox{0.ex}{$1257\pm 9$}
 & \raisebox{0.ex}{$114\pm 14$} & \raisebox{0.ex}{$1318.3\pm 0.6$} & \raisebox{0.ex}{$107\pm 5$} 
\\[-0.ex]
\hline
$f_2^\prime(1525)$ & $1523\pm 3^{+2}_{-8}$ & $71\pm 5^{+17}_{-2}$ &  $1512\pm 3^{+1.4}_{-0.5}$ &  $83\pm  9^{+5}_{-4}$ & $1525\pm 5$ &  $73^{+6}_{-5}$ \\
\hline
 $f_0(1710)$ & $1692\pm 6^{+9}_{-3}$ & $125\pm 12^{+19}_{-32}$ &  $1701\pm 5^{+9}_{-2}$ & $100\pm 24^{+7}_{-22}$ & $1724\pm 7$ & $137\pm 8$  \\
 \hline
\end{tabular}
\caption{\it The measured masses and widths for the $f_2(1270)$, $a_2^0(1320)$,
\ffb and \ffc states using \ksks decays as determined by one fit neglecting interference 
 and another one with interference as predicted by SU(3) symmetry arguments included.
Both statistical and systematic uncertainties are quoted. 
The systematic uncertainty for the  \ffa peak is
expected to be significant and it is not listed.
Also quoted are the PDG values for comparison.
}
\end{center}
\label{tab1}
\end{table}

\newpage

\newpage

\end{document}